\documentclass[a4paper,11pt]{article}
\usepackage{bm,color}
\usepackage{pos}
\usepackage{tikz}
\usepackage{dirtree}
\usepackage{listings}
\usepackage{xcolor}
\usepackage{tikz}
\usepackage{tcolorbox}
\usepackage{fancyvrb}
\usepackage{hyperref}
\usepackage{multirow}
\tcbuselibrary{skins}
\tcbuselibrary{listings}
\tcbuselibrary{breakable}

\newtcblisting{mgscript}[1][]{enhanced,listing only,
boxrule=0.4pt,left=3mm,right=2mm,top=1mm,bottom=1mm,
colback=white,
colframe=black,
boxed title style={size=small,colback=white!70!black,boxrule=0.4pt},
fonttitle=\ttfamily\footnotesize,coltitle=black,
sharp corners,rounded corners=southeast,arc is angular,arc=3mm,
underlay={%
  \path[fill=white!70!black] ([yshift=3mm]interior.south east)--++(-0.4,-0.1)--++(0.1,-0.2);
  \path[draw=black,shorten <=-0.05mm,shorten >=-0.05mm] ([yshift=3mm]interior.south east)--++(-0.4,-0.1)--++(0.1,-0.2);
  },
drop fuzzy shadow,attach boxed title to top right={yshift=-2mm, xshift=-4mm},#1}

\newcommand{\eg}{\emph{e.g.}}
\newcommand{\ie}{\emph{i.e.}}
\newcommand{\cf}{\emph{cf.}}
\newcommand{\mgfull}{\textsc{MadGraph5\_aMC@NLO}}
\newcommand{\mgshort}{\textsc{MG5\_aMC}}
\newcommand{\mdm}{\textsc{MadDM}}
\newcommand{\sttt}[1]{\footnotesize\texttt{#1}} 

\title{Studying dark matter with MadDM~3.1: a short user guide }

\author[a]{Chiara Arina}
\author*[a]{Jan Heisig}
\author[a]{Fabio Maltoni}
\author[a]{Luca Mantani}
\author[a,b]{Daniele Massaro}
\author[a]{Olivier Mattelaer}
\author[c,d]{Gopolang Mohlabeng}

\affiliation[a]{Centre for Cosmology, Particle Physics and Phenomenology (CP3), Universit\'e catholique de Louvain, Chemin du Cyclotron 2, B-1348 Louvain-la-Neuve, Belgium}
\affiliation[b]{Dipartimento di Fisica e Astronomia, Universit{\`a} di Bologna, Viale Berti Pichat 6/2, 40127 Bologna, Italy}
\affiliation[c]{The Arthur B. McDonald Canadian Astroparticle Physics Research Institute and Department of Physics, Engineering Physics and Astronomy, Queens University, Kingston, Ontario, K7L 3N6, Canada}
\affiliation[d]{Perimeter Institute for Theoretical Physics, Waterloo, Ontario, N2L 2Y5, Canada}

\emailAdd{chiara.arina@uclouvain.be}
\emailAdd{jan.heisig@uclouvain.be}
\emailAdd{fabio.maltoni@uclouvain.be}
\emailAdd{daniele.massaro@uclouvain.be}
\emailAdd{olivier.mattelaer@uclouvain.be}
\emailAdd{mohlabeng319@gmail.com}
\emailAdd{luca.mantani@uclouvain.be}

\abstract{\mdm\, is an automated numerical tool for the computation of dark-matter observables for generic new physics models. 
We announce version 3.1 and summarize its features.
Notably, the code goes beyond the mere cross-section computation for direct and indirect detection. For instance, it  allows the user to compute the fully differential nuclear recoil rates as well as the energy spectra of photons, neutrinos and charged cosmic rays for arbitrary $2\to n$ annihilation processes.
This short user guide
equips researchers with all the relevant information required to readily perform comprehensive phenomenological studies of particle dark-matter models.
}

\FullConference{%
  Tools for High Energy Physics and Cosmology - TOOLS2020\\
  2-6 November, 2020\\
  Institut de Physique des 2 Infinis (IP2I), Lyon, France}

\tableofcontents

\begin{document}
\maketitle

\section{Introduction}

The observation of the phenomenon of dark matter on various length scales in our Universe remains one of the major puzzles of modern physics (see \eg~Ref.~\cite{Bertone:2004pz} for a review). The hypothesis of a new particle -- and possibly an entire new sector of particles beyond the standard model (BSM) -- is a widely considered explanation. In particular, the existence of a stable weakly interacting massive particle 
naturally explains the observed value for the relic density via the thermal freeze-out mechanism. While initially being  promoted through the popularity of supersymmetry, by now the idea of frozen-out dark matter has entered the standard recipe for successful dark-matter model building well beyond the supersymmetric paradigm. 
An additional appeal of such a candidate emerges from the promising prospects to detect it. Its production from, scattering off and annihilation into standard-model particles -- probed at colliders, direct and indirect detection experiments, respectively -- provides three complementary search strategies accessible with current experimental sensitivities. 
Exploring the interplay of such observables has become a major direction of phenomenological research and brought forth the need for their efficient numerical computation. This has stimulated the development of automated numerical tools such as \textsc{micrOMEGAs}~\cite{Belanger:2018ccd}, \textsc{DarkSUSY}~\cite{Bringmann:2018lay}, \textsc{SuperIso Relic}~\cite{Arbey:2018msw} and -- in particular, as considered here -- \mdm~\cite{Ambrogi:2018jqj}.

For the computation of cross sections and widths,
\mdm\ utilizes the automatized matrix element generator
\mgfull~\cite{Alwall:2011uj,Alwall:2014hca} (\mgshort\ in the following). It is embedded as a plug-in of the \mgshort\,  platform. As such, \mdm\ supports all particle physics models that can be cast into the Universal FeynRules Output (UFO) format~\cite{Degrande:2011ua}, generated by \eg~\textsc{FeynRules}~\cite{Alloul:2013bka}, \textsc{SARAH}~\cite{Staub:2012pb} or \textsc{LanHEP}~\cite{Semenov:2014rea}.
\mdm~1.0~\cite{Backovic:2013dpa}, released in 2013, introduced the relic density calculator, while versions 2.0~\cite{Backovic:2015cra} and 3.0~\cite{Ambrogi:2018jqj} extended the functionality by a comprehensive set of direct and indirect detection observables, respectively.  
For direct detection, the code not only computes the
elastic spin-independent and spin-dependent dark-matter-nucleon cross sections. It also allows for the computation of the double differential event rates as a function of time, scattering angle and energy supporting a variety of target materials. For indirect detection, \mdm\ allows the user to compute the velocity averaged annihilation cross sections today and the corresponding energy spectra of prompt photons, cosmic rays and neutrinos. The generation of annihilation spectra can either be done by combining pre-computed spectra for individual annihilation channels from \textsc{PPPC4DMID}~\cite{Cirelli:2010xx} (fast mode) or by simulating events employing \textsc{Pythia}~8~\cite{Sjostrand:2014zea} for showering and hadronization (precise mode). The latter enables full flexibility, in particular allowing the user to consider arbitrary $2\to n$ processes. 
The program also includes experimental constraints from the direct detection experiments LUX~\cite{Akerib:2017kat}, Xenon1T~\cite{Aprile:2018dbl} and Pico-60~\cite{Amole:2017dex} as well as indirect detection constraints from gamma-ray observations of dwarf spheroidal galaxies by Fermi-LAT~\cite{Fermi-LAT:2016uux} that allow for the computation of a likelihood and exclusion limit.

With this article, we release version~3.1 that introduces various minor improvements, such as a revised display command, an extended output of the relic density computation as well as updated constraints from Xenon1T~\cite{Aprile:2018dbl}.
We provide a short user guide of \mdm\ that equips researchers with all relevant information required to readily perform comprehensive phenomenological studies of particle dark-matter models. In particular, in section~\ref{sec:getstart} we supply information on the installation and the general functionalities of the code. In section~\ref{sec:DMobs} we detail the observable-specific commands and settings. We summarize and give a brief outlook on upcoming developments in section~\ref{sec:conl}.

\section{Getting started}\label{sec:getstart}

In this section, we provide the basic information on how to install \mdm\, and describe the main commands via a quick tutorial. We depict its folder structure, the relevant output files and give a few tips on how to run it efficiently. 

\subsection{Installation}\label{sec:install}

To install the \mdm\, plug-in, the user has to first download and untar the latest stable version of \mgshort\ from \href{https://launchpad.net/mg5amcnlo}{https://launchpad.net/mg5amcnlo}. At the time of writing, this corresponds to version 2.8.2, which we assume for definiteness in the examples in the following. While \mgshort, is now compatible with  Python 3, this is not yet the case for \mdm, which works only with Python~2.7. Additionally, the user should make sure that there is a complete installation of the SciPy and NumPy modules.\footnote{Note that \mgshort~2.8.X furthermore requires the Python module \texttt{six}.} 
Once the \mgshort\ package has been untared, the user has to enter the corresponding directory, start \mgshort, and install \mdm\ via the \mgshort\ command line:
\begin{verbatim}
    mydir$ tar -xzf MG5_aMC_v2.8.2.tar.gz
    mydir$ cd MG5_aMC_v2_8_2/
    MG5_aMC_v2_8_2$ python2.7 bin/mg5_aMC
    MG5_aMC> install maddm
    MG5_aMC> quit
    MG5_aMC_v2_8_2$
\end{verbatim}
The latest version of \mdm\, will automatically be downloaded and installed as a \mgshort\ plug-in. The corresponding source code in located in \texttt{MG5\_aMC\_v2\_8\_2/PLUGIN/maddm} while the executable python file \texttt{maddm.py} is stored in \texttt{MG5\_aMC\_v2\_8\_2/bin/}. See section~\ref{sec:folder} and figure~\ref{fig:folder} for details on the folder structure.

Note that \mdm\, is automatically interfaced with a few tools that support the computation of indirect-detection observables. 
These are \textsc{Pythia~8}~\cite{Sjostrand:2014zea}, the \textsc{PPPC4DMID} libraries for annihilation spectra~\cite{Cirelli:2010xx}, \textsc{DRAGON}~\cite{Evoli:2008dv} and the \textsc{GALPROP} libraries~\cite{Vladimirov:2010aq} (for \textsc{DRAGON}). 
When performing indirect-detection computations with \mdm\ for the first time, the user is asked whether these packages should be installed automatically. Note that \emph{(i)} the installation can take some time and that \emph{(ii)}~\textsc{Pythia}~8 and \textsc{PPPC4DMID} are needed for the computation of annihilation spectra, while \textsc{DRAGON} is needed for cosmic-ray propagation only. The user can also perform the installation at any time via the \mdm\ command-line interface:
\begin{verbatim}
    MG5_aMC_v2_8_2$ python2.7 bin/maddm.py
    MadDM> install pythia8
    MadDM> install PPPC4DMID
    MadDM> install dragon
    MadDM> install dragon_data_from_galprop
    MadDM> quit
    MG5_aMC_v2_8_2$
\end{verbatim}
For further information about indirect detection and the usage of these packages see section~\ref{sec:ID}.

\subsection{Command-line interface and tutorial mode}\label{sec:basic}

Once the user has entered in \mdm\, by executing the \texttt{maddm.py} file, the first steps for any computation are to load a model and define the dark matter candidate. This is achieved by typing:
\begin{verbatim}
    MG5_aMC_v2_8_2$ python2.7 bin/maddm.py
    MadDM> import model DMsimp_s_spin0
    MadDM> define darkmatter xd
\end{verbatim}
where here we have considered a Dirac dark matter candidate denoted by the particle named \texttt{xd} within the simplified model called \texttt{DMsimp\_s\_spin0}. More details about models are provided in section~\ref{sec:models}.
The relic abundance, direct and indirect detection observables for \texttt{xd} are computed via
\begin{verbatim}
    MadDM> generate relic_density
    MadDM> add direct_detection
    MadDM> add indirect_detection
    MadDM> output my_process_dir
\end{verbatim}

\begin{figure}[b]
\centering
\begin{Verbatim}[fontsize=\footnotesize]
The following switches determine which programs are run:
/============ Description ============|====== values ======|======= other options =======\
| 1. Compute the Relic Density        |     relic = ON     |  OFF                        |
| 2. Compute direct(ional) detection  |    direct = direct |  OFF|directional            |
| 3. Compute indirect detection/flux  |  indirect = sigmav |  flux_source|flux_earth|OFF |
| 4. Run Multinest scan               |  nestscan = OFF    |  ON                         |
\========================================================================================/
 You can also edit the various input card:
 * Enter the name/number to open the editor
 * Enter a path to a file to replace the card
 * Enter set NAME value to change any parameter to the requested value
 /=============================================================================\
 |  5. Edit the model parameters    [param]                                    |
 |  6. Edit the MadDM options       [maddm]                                    |
 \=============================================================================/
 [60s to answer]
 >
\end{Verbatim}
\vspace{-2.3ex}
\caption{Example of the launch interface after performing the \texttt{launch} command in \mdm\@. In the specific example, the relic density, direct and indirect detection calculations are turned on, while the performance of a scan with \textsc{MultiNest} is switched off.
}
\label{fig:lauchpromtp}
\end{figure}

The commands \texttt{generate} and \texttt{add} have the same functionalities they have in \mgshort. In particular, \texttt{generate} is used as the first command, while \texttt{add} is used to retain the previously generated processes and add new ones. Note that a subsequent call of the \texttt{generate} command will erase all previous processes.
The last command above creates a folder \texttt{my\_process\_dir} which contains all the code necessary to launch the required computations. Performing these computations for a given parameter point is done via the \texttt{launch} command.
\begin{verbatim}
    MadDM> launch my_process_dir
\end{verbatim}
This opens the \emph{launch interface} that allows the user to change settings and model parameters, as shown in figure~\ref{fig:lauchpromtp}. There are two ways of making changes. First, (repeatedly) entering a number 1--4 allows to alternate between the options displayed, while entering 5 or 6 opens the files \texttt{param\_card.dat} or \texttt{maddm\_card.dat}, respectively, with a command-line editor (\texttt{vim} by default).\footnote{Note that the files may, of course, be changed by any other instance instead.} These files contain all model parameters and most of the \mdm\ settings, respectively. A second option is to directly type \texttt{\,set\;<parameter>\;<value>\,} in the launch interface,~for instance
\begin{verbatim}
    > set mxd 500
\end{verbatim}
for setting the dark-matter mass to 500\,GeV. Auto-completion is available (via pressing tab) to easily find the name of parameters. The observable-specific settings (the first four entries of figure~\ref{fig:lauchpromtp}) will be described in detail in section~\ref{sec:DMobs}.
Once the user is done with all settings the launch interface is finally exited by pressing enter.

Note that the \texttt{launch} command can be executed either in the same session
or after quitting and restarting \mdm\@. In the former case, the specification of the directory where the process has been created is not necessary, as \mdm\ will launch the process of the last output in the session.  

\medskip

A convenient way of being guided through the basic commands is the tutorial model. It is entered by typing \texttt{tutorial} in the \mdm\  command-line interface:
\begin{verbatim}
    MG5_aMC_v2_8_2$ python2.7 bin/maddm.py
    MadDM> tutorial
\end{verbatim}
The screen output explains the basic commands and options that the user may follow. It can be exited by:
\begin{verbatim}
    MadDM> tutorial stop 
\end{verbatim}

\subsection{The \texttt{display} command}

When computing the observables for dark matter models, it is possible to use the following commands
\begin{verbatim}
    MadDM> display processes
    MadDM> display diagrams
\end{verbatim}
to either display a list of the generated processes or the respective Feynman diagrams.
From \mdm~3.1 on, the display command allows for the following options:
\begin{itemize}
    \item \texttt{relic}, \texttt{direct} or \texttt{indirect}: display only processes/diagrams related to relic density, direct detection or indirect detection; a combination of them is supported, see the example below;
    \item \texttt{last}: displays only processes/diagrams generated by the last command called, it overwrites any other option specified;
    \item \texttt{all}: works as the simultaneous presence of \texttt{relic}, \texttt{direct}, \texttt{indirect}: it displays only diagrams relevant for dark matter annihilation and it is the default setting if no options are provided.
\end{itemize}
For instance, the command
\begin{verbatim}
    MadDM> display processes relic indirect
\end{verbatim}
displays all the processes related to relic density and indirect detection.

\subsection{Running \texorpdfstring{\mdm}{MadDM} from a script}\label{sec:script}

In certain applications, it might not be convenient or even possible to use the command-line interface of \mdm\ described in section~\ref{sec:basic}.
An alternative is to control \mdm\ via a script.
To do so the respective commands described in section~\ref{sec:basic} need simply to be written in a plain text file separated by line-breaks. The respective script can be passed as an argument when starting \mdm. The corresponding operations will then be executed. For instance, the user may create the two scripts:\footnote{Note that the content of the two scripts could as well be put into one file. The separation of the launch command is, however, often convenient as only this part needs to be rerun when choosing different parameters.}\medskip\\
\noindent
\begin{mgscript}[adjusted title=generate.txt,width=0.45\linewidth,nobeforeafter,box align=top]
import model DMsimp_s_spin0
define darkmatter xd
generate relic_density
add direct_detection
add indirect_detection
output my_process_dir
\end{mgscript}
\hfill
\begin{mgscript}[adjusted title=launch.txt,width=0.45\linewidth,nobeforeafter,box align=top]
launch my_process_dir
indirect = flux_source
direct = direct
set sigmav_method madevent
set nevents 20000
set mxd 500
\end{mgscript}
\vspace{3ex}

\noindent
and execute MadDM:
\begin{verbatim}
    MG5_aMC_v2_8_2$ python2.7 bin/maddm.py generate.txt
    MG5_aMC_v2_8_2$ python2.7 bin/maddm.py launch.txt
\end{verbatim}
Note that the settings \texttt{relic}, \texttt{direct}, \texttt{indirect} and \texttt{nestscan} should not be set by
entering the numbers 1--4 (\cf~section~\ref{sec:basic}) in scripts as the selected mode can depend on the machine on which the code runs and on the specific \mdm\,  version installed.
The observable-specific commands contained in \texttt{launch.txt} are detailed in section~\ref{sec:DMobs}.

\subsection{Folder structure}\label{sec:folder}

\begin{figure}[t]
    \centering
    {
    \begin{minipage}{6cm}
        \dirtree{%
        .1 MG5\_aMC\_v2\_8\_2. 
        .2 bin. 
        .3 maddm.py.
        .3 mg5\_aMC.
        .3 \dots.
        .2 models.
        .2 my\_process\_dir.
        .3 bin.
        .3 Cards.
        .4 maddm\_card.dat.
        .4 multinest\_card.dat.
        .4 param\_card.dat.
        .4 \dots.
        .3 Indirect.
        .3 output.
        .4 run\_01.
        .4 run\_02.
        .4 \dots.
        .3 \dots.
        .2 PLUGIN.
        .3 maddm.
        .3 \dots.
        .2 \dots.
        }
    \end{minipage}
        \begin{minipage}{8.5cm}
        \dirtree{%
        .1 run\_01.
        .2 maddm\_card.dat.
        .2 MadDM\_results.txt.
        .2 maddm.out.
        .2 Output\_Indirect.
        .3 antiprotons\_spectrum\_pythia8.dat.
        .3 gammas\_spectrum\_pythia8.dat.
        .3 neutrinos\_e\_spectrum\_pythia8.dat.
        .3 neutrinos\_mu\_spectrum\_pythia8.dat.
        .3 neutrinos\_tau\_spectrum\_pythia8.dat.
        .3 positrons\_spectrum\_pythia8.dat.
        .3 restx\_spectrum\_pythia8.dat.
        .3 pythia8.log.
        .3 run\_01\_DM\_banner.txt.
        .3 run\_shower.sh.
        .3 unweighted\_events.lhe.gz.
        }
    \end{minipage}
    }
    \caption{Schematic structure of the folders and files of \mgshort\, and \mdm. {\bfseries{On the left:}}  Main directory of \mgshort\, where the python executable file \texttt{maddm.py} is located in the \texttt{bin} folder, while the source code is in \texttt{PLUGIN/maddm/}. The output directory \texttt{my\_process\_dir} contains all relevant setting cards (within \texttt{Cards}), and the output files in \texttt{output/run\_01} for instance. {\bfseries{On the right:}} Zoomed view of the \texttt{run\_01} directory, where the main results are stored, as labeled. The file \texttt{MadDM\_results.txt} recaps the value of all observable computed by the user. Notice that \texttt{Output\_Indirect} contains indirect-detection files, such as the energy spectra and the lhe event file.}
    \label{fig:folder}
\end{figure}

Figure~\ref{fig:folder} displays the general folder structure of \mgshort\ after installation of the \mdm\,  plug-in. The directory \texttt{bin} contains the python code to be executed, while \texttt{models} contains all models used, see section~\ref{sec:models} for more information. Once generated, \texttt{my\_process\_dir} contains all code, input and output for a certain process. Input parameters are stored in various files in the directory \texttt{Cards}. For instance, model parameters can be either set via the \texttt{set} command in the launch interface (see section~\ref{sec:basic}) or by changing the respective parameters in \texttt{Cards/param\_card.dat}. The output folder contains a sub-directory for each run within the process \texttt{run\_01}, \texttt{run\_02}, \dots which, in turn, contains all outputs of the computation (stored in \texttt{MadDM\_results.txt} and \texttt{maddm.out}) as well as a copy of the \texttt{maddm\_card.dat} used. Further output is linked to the directory \texttt{Output\_Indirect}, see section~\ref{sec:ID} for more details.

\subsection{Running scans}\label{sec:scans}

There are several ways to run scans over parameter space points within \mdm. First, \mdm\,  may be just called by an external code that performs a scan. In this case, the parameters and settings may just be passed by a script as detailed in section~\ref{sec:script}.

The second option is to employ the sequential grid scan functionality of \mdm, which allows one to scan over an arbitrary number of model parameters with one launch command. To achieve this, instead of setting a given parameter to a fixed value, the respective scan range has to be defined:
\begin{verbatim}
    MadDM> launch my_process_dir
    > set mxd scan:range(50,700,25)
\end{verbatim}
Note that after the syntax \texttt{scan:} any python iterable is accepted, including list comprehension syntax. In the case of multi-dimensional scans, two possibilities are available. Using the syntax \texttt{scan:} for two or more parameters will generate a nested loop over the scan ranges, i.e. a complete grid. Another possibility is to use the syntax \texttt{scan1:}, which instead creates a parallel scan, namely the values of the iterables are scanned simultaneously.
Instead of being specified in the launch interface, this setting can also be done in the \texttt{param\_card.dat}:
\begin{center}
\begin{mgscript}[title=my\_process\_dir/Cards/param\_card.dat, width=9cm]
...
Block mass
    1 5.040000e-03 # MD 
    2 2.550000e-03 # MU 
    3 1.010000e-01 # MS 
    4 1.270000e+00 # MC 
    5 4.700000e+00 # MB 
    6 1.720000e+02 # MT 
   15 1.777000e+00 # MTA 
   23 9.118760e+01 # MZ 
   25 1.250000e+02 # MH 
   51 1.000000e+01 # MXc 
   52 scan:range(50,700,25) # MXd 
   54 1.000000e+03 # MY0 
   ...
\end{mgscript}
\end{center}

The third option is to perform guided scans with the Bayesian inference tool \textsc{MultiNest} (\textsc{MultiNest}~\cite{Feroz:2007kg,Feroz:2008xx} is provided together with \textsc{PyMultiNest}~\cite{Buchner:2014nha}) by specifying
\begin{verbatim}
    MadDM> launch my_process_dir
    > nestscan = 0N
\end{verbatim}
and setting multinest parameters in \texttt{multinest\_card.dat}, which then appears as number 7 in the launch interface:
\begin{Verbatim}[fontsize=\footnotesize]
   /=================================================================\
   |  5. Edit the model parameters    [param]                        |
   |  6. Edit the MadDM options       [maddm]                        |
   |  7. Edit the Multinest options   [multinest]                    |
   \=================================================================/
   [60s to answer]
   >
\end{Verbatim}
For further details see Appendix~E.2 of~Ref.~\cite{Ambrogi:2018jqj}.

\subsection{Importing models}\label{sec:models}

Being a plug-in of \mgshort\,, \mdm\, can perform computation within any particle physics model that allows for an implementation in the Universal FeynRules Output (UFO) format~\cite{Degrande:2011ua}. The implementation can be achieved with automated tools like \textsc{FeynRules}~\cite{Alloul:2013bka}, \textsc{SARAH}~\cite{Staub:2012pb} or \textsc{LanHEP}~\cite{Semenov:2014rea}.
The respective model directory (named as the model to be imported) has to be stored in the directory \texttt{models} (\cf~section~\ref{sec:folder}). Note that a database of models can be found at the FeynRules webpage: \href{http://feynrules.irmp.ucl.ac.be/wiki/ModelDatabaseMainPage}{http://feynrules.irmp.ucl.ac.be/wiki/ModelDatabaseMainPage}. For models stored in that database (but not in the local \texttt{models} folder) \mdm\, automatically downloads the model importing it via the \mdm\, command-line interface (\cf~section~\ref{sec:basic}). This model list can be viewed by the user by typing 
\begin{verbatim}
    MadDM> display modellist
\end{verbatim}
The entire list of models is then displayed.
A user guide for the implementation of a particle physics model into \textsc{FeynRules} can be found in Ref.~\cite{Alloul:2013bka}.

\section{Dark matter observables}\label{sec:DMobs}

In the following, we detail the capabilities of \mdm\,  to perform computations of the relic density, direct and indirect detection observables, respectively.

\subsection{Relic density}\label{sec:reldens}

\mdm\,  allows for the computation of the relic density in the framework of thermal freeze-out of dark matter. It automatically computes the rates for all relevant $2\to2$ annihilation processes including coannihilation processes. 
Coannihilation is taken into account if the user specifies the coannihilating partner(s), such as 
\begin{verbatim}
    MadDM> define coannihilator xco1 xco2 xco3
\end{verbatim}
prior to the command \texttt{generate relic\_density}, see section~\ref{sec:basic}. Here \texttt{xco1}, \texttt{xco2}, \texttt{xco3} are exemplary names of coannihilating partners in the model.
The code solves the corresponding Boltzmann equation for the dark-matter abundance (or dark-sector\footnote{Here we consider the \emph{dark sector} to comprise the dark matter candidate and all potential coannihilators.} abundance for the case of coannihilation~\cite{Edsjo:1997bg}) numerically, see Ref.~\cite{Backovic:2013dpa} for further details. It assumes kinetic equilibrium between all involved particles (as well as chemical equilibrium within the dark sector) during dark-matter freeze out.

The output on screen is, for instance:
\begin{verbatim}
    ***** Relic Density
    OMEGA IS  0.000325869586293
    INFO: Relic Density = 3.26e-04 UNDERABUNDANT
    INFO: x_f           = 2.80e+01
    INFO: sigmav(xf)    = 3.27e-24 cm^3/s
    INFO: xsi           = 2.72e-03
\end{verbatim}
The relic density is given in terms of $\Omega h^2$, where $\Omega$ is the dark-matter energy density in units of the critical density and $h$ is the Hubble constant in units of $100\,\text{km}\,\text{s}^{-1}\, \text{Mpc}^{-1}$.
The freeze-out point (approximately the point of chemical decoupling) is given by $x_\text{f}$, where $x =m_\text{DM}/T$. 
The thermally averaged annihilation cross section at the freeze-out point, $\langle\sigma v\rangle (x_\text{f})$, is given in units of $\text{cm}^{3}\,\text{s}^{-1}$.
Since version 3.1, \mdm\ additionally displays the contributions of the different channels to the relic density.
The resulting screen output reads:
\begin{verbatim}
    INFO: Channels contributions:
    INFO: xdxdx_hh            : 0.05 %
    INFO: xdxdx_zz            : 0.13 %
    INFO: xdxdx_ttx           : 98.21 %
    INFO: xdxdx_aa            : 0.95 %
    INFO: xdxdx_wpwm          : 0.10 %
    INFO: xdxdx_az            : 0.57 %
    INFO: No contribution from processes: y0y0
\end{verbatim}
Models whose relic density undershoots (overshoots) the value measured by Planck, $\Omega h^2 = 0.120\pm 0.001$~\cite{Aghanim:2018eyx}, by more than $2\sigma$ are flagged as \texttt{UNDERABUNDANT} (\texttt{OVERABUNDANT}), while values within that range are flagged as \texttt{WITHIN\_EXP\_ERRORS}. However, note that the theoretical uncertainty in the model prediction -- typically significantly larger -- is not estimated by \mdm. 
It is up to the user to estimate the error on $\Omega h^2$ to be taken into account \eg~in a global fit.

In the case of thermally underabundant dark matter, \texttt{xsi} denotes the fraction of the model's thermal abundance of total measured dark matter abundance,
\begin{equation}
    \xi= \frac{(\Omega h^2)_\text{model}}{(\Omega h^2)_\text{Planck}}.
\end{equation}
In this case constraints from direct and indirect detection, subject to the following sections, are interpreted in two ways:
\begin{itemize}
    \item
One assumes that the model's candidate, indeed only makes up a fraction $\xi$ of the total amount of dark matter, implying the existence of a further unspecified contribution, \eg~axions or primordial black holes. This interpretation entails the rescaling of the yields for direct and indirect detection by a factor of $\xi$ and $\xi^2$, respectively. It is denoted by `Thermal'. 
    \item
Regardless of the underabundant contribution from thermal freeze-out, the model's candidate is assumed to constitute 100\% of the measured dark matter abundance. This interpretation applies in the presence of an additional (non-thermal) contribution to dark-matter production, \eg~through a late decay of a heavier species. It is denoted by `All DM'. 
\end{itemize}
Note that to enable the interpretation of direct and indirect detection observables in the `Thermal' scenario the relic density computation has to be performed.

\subsection{Direct detection}\label{sec:DD}

Direct detection experiments search for dark-matter particles scattering off atomic nuclei in low-background environments, \eg~deep underground. If the recoil momentum of the nucleus is above the detection threshold, then electrons, photons and/or phonons induced by the nuclear recoil may be detected. The number of dark-matter scattering events in a given experiment is then set by a confluence of factors in the dark-matter theory parameter space.
This includes the mass, scattering cross section and the astrophysical velocity distribution of the dark matter in our local Galactic neighborhood. 
Since the solar system is moving in a particular direction with respect to the Galactic dark-matter halo, the astrophysical  distribution can provide both velocity and angular information. 
These are used to calculate the nuclear recoil energy spectrum and the angular recoil spectrum, by both direct detection and directional detection experiments respectively.

\mdm\ allows the user to choose among two modes called \texttt{direct} and \texttt{directional}. The setting can be changed in the launch interface either by (repeatedly) entering the number 2 until the requested option is displayed on screen, \cf~figure~\ref{fig:lauchpromtp} and section~\ref{sec:basic}, or by directly entering one of the following commands:
\begin{verbatim}
    > direct = direct
    > direct = directional
\end{verbatim}

The mode \texttt{direct} provides
the basic computations of spin-independent and spin-dependent dark-matter-nucleon cross section as well as their respective limits from LUX~\cite{Akerib:2017kat}, XENON1T~\cite{Aprile:2018dbl}, and PICO-60~\cite{Amole:2017dex}. An exemplary screen output from the direct detection module (for the same parameters used above) reads:
\begin{Verbatim}[fontsize=\scriptsize]
    ***** Direct detection [cm^2]:
    INFO: SigmaN_SI_p: Thermal = 2.01e-50 ALLOWED, All DM = 7.40e-48 ALLOWED  Xenon1ton ul = 4.16e-46
    INFO: SigmaN_SI_n: Thermal = 1.98e-50 ALLOWED, All DM = 7.27e-48 ALLOWED  Xenon1ton ul = 4.16e-46
    INFO: SigmaN_SD_p: Thermal = 0.00e+00 ALLOWED, All DM = 0.00e+00 ALLOWED  Pico60 ul    = 2.03e-40
    INFO: SigmaN_SD_n: Thermal = 0.00e+00 ALLOWED, All DM = 0.00e+00 ALLOWED  Lux2017 ul   = 1.22e-40
\end{Verbatim}
As indicated, all direct-detection cross sections are given in $\text{cm}^2$. In the `Thermal' scenario the cross-section prediction is rescaled by $\xi$, see section~\ref{sec:reldens} for details. This particular model does not have spin-dependent interactions due to the type of mediator interacting with the dark matter.

The mode \texttt{directional} additionally provides the fully differential nuclear recoil rates as a function of energy, angle and time. 
It, therefore, allows the user to explore the directional information of dark-matter scattering.
For the computation of nuclear recoil rates
\mdm\ allows the user to choose among a large set of detector materials and take into account detector smearing effects. Furthermore, the recoil energy, the detector size, the most probable dark-matter velocity and escape velocity as well as the local dark-matter density can be specified. Moreover, the nuclear form factor can be customized. All these settings can be adjusted by changing the corresponding entries in the \texttt{maddm\_card.dat} file. For more information, see~\cite{Backovic:2015cra}.

\subsection{Indirect detection}\label{sec:ID}

Indirect detection probes the (self-)annihilation of dark matter in locally over-dense regions, like the center of the Galaxy. Stable particles that are the final products of these annihilation processes can propagate to us and act as the messengers of the dark-matter signal. 
Photons (gamma rays), neutrinos and stable charged particles (cosmic rays), in particular, antiprotons and positrons, are commonly considered messenger particles. The dark-matter annihilation cross section and energy spectra of these messengers need to be computed to confront the signal prediction with data. 
\mdm\ provides two different modes and a variety of further settings to supply the user with these observables at an appropriate level of precision and speed. 

\subsubsection{Running modes and settings}

In the \texttt{fast} mode, the cross-section computation is performed with a fast phase-space integrator using the Simpson method~\cite{Weinzierl:2000wd} (also used for the relic density computations). In this mode, no events are generated. Furthermore, it is restricted to $2\to2$ processes. It is selected by entering, in the \texttt{launch} interface, the command 
\begin{verbatim}
    > set fast
\end{verbatim}

In \texttt{precise} mode, the phase-space integration is performed by \textsc{MadEvent}~\cite{Maltoni:2002qb}. Events are generated and arbitrary $2\to n$ processes can be taken into account. 
It is selected by
\begin{verbatim}
    > set precise
\end{verbatim}
The \texttt{precise} mode allows the user to evaluate the cross section either at a fixed dark-matter velocity 
\begin{verbatim}
    > set sigmav_method = madevent
\end{verbatim}
or taking into account a Maxwell-Boltzmann distribution in velocity through a reshuffling and reweighting of events~\cite{Kleiss:1985gy,Mattelaer:2016gcx}:
\begin{verbatim}
    > set sigmav_method = reshuffling
\end{verbatim}
Notice that this is the default setting if nothing is specified.
The average velocity (in units $c=1$) can be set as follows, \eg~to $10^{-5}$:
\begin{verbatim}
    > set vave_indirect 1e-5
\end{verbatim}
which is the default value if nothing is specified. Note that the automatic derivation of constraints from dwarf spheroidal galaxies (see below) requires the velocity to lie between $1.4\times10^{-4}$ and $3\times10^{-6}$, while typical velocities for the Galactic center is around $10^{-3}$.
The number of generated events can be set as follows, for instance:
\begin{verbatim}
    > set nevents = 50000
\end{verbatim}
Note that the generation of smooth spectra (in particular towards the tails) might require a large number of events (up to several million). However, for analyses of binned spectra much fewer events can often be sufficient. For instance, for the computation of Fermi-LAT limits (see below), based on a binned likelihood function (with 24 bins) an event number between 10000 to 50000 is often sufficient to obtain a good estimate of the constraints. However, as the number and energy of messenger particles per annihilation depends strongly on the dark-matter model, these numbers are not universally valid. For instance, considering heavy dark matter with masses larger than a few TeV, Fermi-LAT only probes the low-energy tail of the photon spectrum that is sampled by a small fraction of events. In such cases, care has to be taken when estimating the number of required events.

In both modes, \texttt{fast} and \texttt{precise}, the user can specify whether to compute just the cross section (\texttt{sigmav}) or in addition the energy spectra at sources (\texttt{flux\_source}) or near Earth (\texttt{flux\_earth}). The latter is relevant for neutrinos, which oscillate, and for cosmic rays that are subject to a non-trivial propagation process between the source and Earth that, in particular, affect the spectra. This setting can be chosen by repeatedly typing the number 3 in the launch interface to alternate between the three options (and \texttt{OFF}), \cf~figure~\ref{fig:lauchpromtp} and section~\ref{sec:basic}. Alternatively, it can be set by one of the following commands, respectively:
\begin{verbatim}
    > indirect = sigmav
    > indirect = flux_source
    > indirect = flux_earth
\end{verbatim}
This provides a total of six different running modes. The respective default settings and further options are summarized in table~\ref{tab:options}. We will briefly discuss them in the following for completeness, see~\cite{Ambrogi:2018jqj} for further details.

\begin{table}[h!]
\begin{center}
\begin{tabular}{|c|l|l|}
\hline

& \texttt{fast} mode  & \texttt{precise} mode \\
\hline
\multirow{8}{*}{\rotatebox[origin=c]{90}{\texttt{indirect\;=\;sigmav}}}&   & \\
& Default:  & Default:\\
& \sttt{sigmav\_method\;=\;inclusive}  & \sttt{sigmav\_method\;=\;reshuffling}\\
&    & \\
&    & Other options:\\
&    & {\sttt{sigmav\_method\;=\;madevent}}\\
&    & \\
&    & \\
\hline
\multirow{10}{*}{\rotatebox[origin=c]{90}{\texttt{indirect\;=\;flux\_source}}}&   & \\
& Default:  & Default:\\
& {\sttt{sigmav\_method\;=\;inclusive}}  & {\sttt{sigmav\_method\;=\;reshuffling}}\\
& {\sttt{indirect\_flux\_source\_method\,=\,PPPC4DMID\_ew}} &{\sttt{indirect\_flux\_source\_method\;=\;pythia8}} \\
&   & \\
&  Other options:  & Other options: \\
&   {\sttt{indirect\_flux\_source\_method\,=\,PPPC4DMID}} & \sttt{sigmav\_method\;=\;madevent}\\
&    &{\sttt{indirect\_flux\_source\_method\,=\,PPPC4DMID\_ew}} \\
&    & {\sttt{indirect\_flux\_source\_method\,=\,PPPC4DMID}}  \\
&    & \\
\hline
\multirow{10}{*}{\rotatebox[origin=c]{90}{\texttt{indirect\;=\;flux\_earth}}}&   & \\
& Default:  & Default:\\
& {\sttt{sigmav\_method\;=\;inclusive}}  & {\sttt{sigmav\_method\;=\;reshuffling}}\\
& {\sttt{indirect\_flux\_earth\_method\,=\,PPPC4DMID\_ep}} &{\sttt{indirect\_flux\_source\_method\;=\;pythia8}} \\
&   & {\sttt{indirect\_flux\_earth\_method\;=\;dragon}} \\
&    & \\
&   & Other options: \\
&   & \sttt{sigmav\_method\;=\;madevent}\\
&    &{\sttt{indirect\_flux\_earth\_method\,=\,PPPC4DMID\_ep}} \\
&    & \\
\hline
\end{tabular}
\end{center}
\caption{Summary of the \mdm\ indirect-detection functionalities upon the execution of the launch command. We display the default settings and further options of all six occurring combinations.}
\label{tab:options}
\end{table}

In \texttt{fast} mode, enabling \texttt{indirect\;=\;flux\_source}, the energy spectra are taken from the \textsc{PPPC4DMID} database~\cite{Cirelli:2010xx} containing pre-computed results for annihilation into pairs of standard model particles only. \mdm\ combines the spectra of different channels according to their cross sections.
The user may choose whether to include electroweak corrections~\cite{Ciafaloni:2010ti} (default) or not, specified by setting \texttt{indirect\_flux\_source\_method} to \texttt{PPPC4DMID\_ew} or \texttt{PPPC4DMID}, respectively.
Note that in \texttt{fast} mode, propagated cosmic-ray spectra are only available for positrons.

In \texttt{precise} mode, enabling \texttt{indirect\;=\;flux\_source}, energy spectra are computed using the generated events and \textsc{Pythia}~8~\cite{Sjostrand:2014zea} for showering and hadronization, where electroweak corrections are enabled by default.\footnote{This setting can be changed by modifying \texttt{pythia\_card.dat}.}
However, the user can also choose to use the pre-computed spectra from \textsc{PPPC4DMID} instead (\cf~table~\ref{tab:options}) by using the commands mentioned above. 
For \texttt{indirect\;=\;flux\_earth} the energy spectra of charged particles are propagated by the numerical code \textsc{DRAGON}~\cite{Evoli:2008dv}.
The propagation parameters can be set in the \texttt{dragon\_card.xml}.
Similar to the case of source spectra, for positrons, the propagates spectra can alternatively be taken from the \textsc{PPPC4DMID} database, regardless of the \texttt{precise} mode. This is achieved by 
typing:
\begin{verbatim}
    > indirect_flux_earth_method = PPPC4DMID_ep
\end{verbatim}
Note that the propagated neutrino spectra are always computed after oscillations in vacuum employing the very long baseline approximation (see~\cite{Ambrogi:2018jqj} for details).

As explained in section~\ref{sec:install}, both \textsc{Pythia}~8 and \textsc{DRAGON} are automatically installed within the \mdm\ framework 
either when first running \texttt{indirect\_detection} (asked for in the launch interface) or via the command-line interface whenever the user needs them.
We ask the user to cite these additional public codes when these are used within \mdm.

\subsubsection{Selecting processes}

So far we have assumed that the user employs the general command (\cf~section~\ref{sec:basic})
\begin{verbatim}
    MadDM > generate indirect_detection
\end{verbatim}
to specify the considered processes.
Note that this command generates all processes where two dark matter particles annihilate into two particles, including all possible standard model particles and BSM particles which are even under the `dark' group.
Alternatively, the user can specify the final state by typing:
\begin{verbatim}
    MadDM > generate indirect_detection u u~
\end{verbatim}
forcing \mdm\ to generate the diagram with a pair of $u$-quarks in the final state only. In fact, any $2\to n$ dark-matter annihilation process can be computed (requiring running \texttt{precise} mode). For instance, one may consider internal bremsstrahlung, where an additional photon \texttt{a} is emitted:
\begin{verbatim}
    MadDM > generate indirect_detection u u~ a
\end{verbatim}
Note that further \mgshort\ syntax can be employed to specify the process. For instance,
to collectively specify a set of particles, multiparticle variables may be defined:
\begin{verbatim}
    MadDM > define q = u d s c b t
    MadDM > define qbar =  u~ d~ s~ c~ b~ t~
\end{verbatim}
Furthermore, the decay of final state particles may be specified (and hence performed by \mdm). For example, the model used above as reference contains a spin-0 mediator \texttt{y0} that can appear in the final state (if it is lighter than dark matter). The mediator can further decay into quarks. Assuming the above multiparticle definition we can hence specify its appearance in the final state and subsequent decay by
\begin{verbatim}
    MadDM > generate indirect_detection y0 y0, y0 > q qbar
\end{verbatim}

Alternatively, decays can be performed by \textsc{Pythia}~8 if the particles' branching ratios are provided. To this end, an automatic computation of branching ratios within \mdm\ can be performed by setting the corresponding decay width to \texttt{AUTO} in the \texttt{param\_card.dat}~\cite{Alwall:2014bza}. The above process can, hence, also be computed by typing:
\begin{verbatim}
    MadDM > generate indirect_detection y0 y0
\end{verbatim}
while the decay width is set in the launch interface:
\begin{verbatim}
    > set wy0 AUTO
\end{verbatim}
(or by the corresponding change in the \texttt{param\_card.dat} by any other instance). With these settings,
\mdm\ automatically computes all branching ratios of the spin-0 mediator while \textsc{Pythia}~8 performs the respective decays in the narrow width approximation.

\subsubsection{Fermi-LAT constraints}

Once the photon energy spectra have been computed with one of the methods described above, \mdm\, automatically computes the exclusion limit from the Fermi-LAT gamma-ray data from dwarf spheroidal galaxies~\cite{Fermi-LAT:2016uux}, if the dark matter velocity is set to an allowed value, \ie~between $1.4\times10^{-4}$ and $3\times10^{-6}$. For details on the Fermi-LAT likelihood function implementation, we refer to~\cite{Ambrogi:2018jqj}. The output is given by the excluded annihilation cross section at 95\% CL (confidence level) compared to the predicted annihilation cross-section, the Fermi-LAT likelihood and the $p$-value for the tested model point. 

\subsubsection{Output}

An exemplary screen output from the indirect detection module (for the same parameters used above) reads:
\begin{Verbatim}[fontsize=\scriptsize]
****** Indirect detection [cm^3/s]:
INFO: <sigma v> method: madevent
INFO: DM particle halo velocity: 2e-05/c
INFO: xdxdx_zz      Thermal = 1.67e-38  ALLOWED     All DM = 2.26e-33  ALLOWED     Fermi ul = 9.04e-23
INFO: xdxdx_aa      Thermal = 1.24e-37  NO LIMIT    All DM = 1.69e-32  NO LIMIT    Fermi ul = -1.00e+00
INFO: xdxdx_ttx     Thermal = 1.28e-35  ALLOWED     All DM = 1.74e-30  ALLOWED     Fermi ul = 1.11e-25
INFO: xdxdx_hh      Thermal = 6.01e-39  ALLOWED     All DM = 8.15e-34  ALLOWED     Fermi ul = 2.21e-22
INFO: xdxdx_wpwm    Thermal = 1.27e-38  ALLOWED     All DM = 1.73e-33  ALLOWED     Fermi ul = 1.16e-22
INFO: Skipping zero cross section processes for: xrxr, xcxcx, y0y0
INFO: Total limits calculated with Fermi likelihood:
INFO: DM DM > all   Thermal = 1.30e-35  ALLOWED     All DM = 1.76e-30  ALLOWED     Fermi ul = 2.85e-25
INFO:
INFO: *** Fluxes at earth [particle/(cm^2 sr)]:
INFO: gammas Flux         =     1.57e-14
INFO: neutrinos_e Flux    =     8.89e-18
INFO: neutrinos_mu Flux   =     9.71e-18
INFO: neutrinos_tau Flux  =     8.75e-18
\end{Verbatim}
For each annihilation channel, the velocity averaged annihilation cross section today $\langle \sigma v \rangle_0$ is displayed in units of $\text{cm}^3 \text{s}^{-1}$. For the `Thermal' scenario, the cross section predictions are rescaled by $\xi^2$, see section~\ref{sec:reldens} for details.
Processes with zero cross section are listed below.
Each channel is compared to the limit coming from the Fermi-LAT constraints on prompt photons from dwarf spheroidal galaxies.
Subsequently, the total cross section is shown, with the limit computed performing the full Fermi-LAT likelihood analysis. 
The last lines show the values of the total integrated flux for prompt photons and neutrinos.

In addition, several output files are produced.
In the case of single point runs, the relevant output is written into \texttt{my\_process\_dir/output/run\_01/MadDM\_results.txt} (\cf~figure~\ref{fig:folder}). The file contains a summary of the computed observables, including the Fermi-LAT likelihood and $p$-value for the considered parameter point. It is formatted conveniently to enable parsing.
The output spectra generated from the PPPC4DMID database can be found in the same directory. The ones generated by using simulated events (in the \texttt{precise} mode) are stored in the \texttt{my\_process\_dir/Indirect/Events/run\_01/} directory, \cf~the folder structure shown on the right in figure~\ref{fig:folder}.
In case of a scan, \texttt{MadDM\_results.txt} is not created. Instead, the file \texttt{my\_process\_dir/output/scan\_run\_01.txt} is written.
It contains a list of the computed observables for all parameter points scanned over.

\section{Summary and outlook}\label{sec:conl}

\mdm\,  is a comprehensive numerical tool for performing computations of dark-matter observables. 
In particular, it supports a detailed interpretation of direct and indirect dark-matter searches by providing \eg~the fully differential nuclear recoil rates (as a function of energy, angle and time) as well as the photon, neutrino and cosmic-ray spectra for arbitrary $2\to n$ annihilation processes at source or near Earth. For the latter, \mdm\ is interfaced to \textsc{Pythia}~8 and \textsc{Dragon}.

Being a plug-in of \mgshort, \mdm\ is conveniently installed and run through a user-friendly command-line interface. It provides an interactive and self-explanatory tutorial mode while the experienced user may prefer running \mdm\ via scripting.
The \mgshort\ framework provides further features  inherited by \mdm\ such as automated width computation or natural support of any particle physics model that can be cast in a UFO format.

\mdm\ is subject to ongoing developments that further enlarge its capabilities.
For the next release, the computations are extended to general loop-induced processes.
In particular, we will provide a framework to analyze the gamma-ray line spectrum arising from the annihilation of dark matter into photons, like $\gamma\gamma$, $\gamma Z$, $\gamma h$.
An automated computation of 
constraints from the gamma-ray line searches from observations of the Galactic center~\cite{Ackermann:2015lka,Abdallah:2018qtu} will also be supplied.

\section*{Acknowledgements}

C.A.~is supported by the Innoviris  ATTRACT 2018 104 BECAP 2 agreement. 
J.H.~acknowledges support from the F.R.S.-FNRS, of which he is a postdoctoral researcher. This work has received funding 
from the European Union's Horizon 2020 research and innovation program 
as part of the Marie Sk\l{}odowska-Curie Innovative Training Network MCnetITN3
(grant agreement no.~722104). Computational resources have been provided by the supercomputing facilities of the Universit{\'e} catholique de Louvain (CISM/UCL) and the Consortium des {\'E}quipements de Calcul Intensif en F{\'e}d{\'e}ration Wallonie Bruxelles (C{\'E}CI) funded by the Fond de la Recherche Scientifique de Belgique (F.R.S.-FNRS) under convention 2.5020.11 and by the Walloon Region.

\bibliographystyle{JHEP}
\bibliography{biblio}

\end{document}